# 3T-1R Analog Write and Digital Read of MRAM for RNG and Low Power Memory Application

Thomas Egler[1,2], Hans Dittmann[1,2], Sunanda Thunder[1] and Artur Useinov[1*]

[1]National Chiao Tung University, 1001 University Rd., Hsinchu 30010, Taiwan, *Email: artu@nctu.edu.tw
[2]Reutlingen University, 50 Oferdinger Str., Reutlingen 72768, Germany

*Abstract* — This work represents integration of MTJ with 30nm FinFET for low voltage analog write operations and readout optimization for the p-bit or true random number generator (TRNG), where the induced p-bit, the probabilistic state of the magnetic tunnel junction (MTJ), is detected within only a single computational period. The period contains two sub-cycles: write and joined read & reset cycles. The operation with MTJ becomes stochastic, independent after calibrating at the desired working point against the factors, which can induce the signal deviations, e.g. temperature, material degradation or external magnetic field.

## I. Introduction

In the present work, we improve and reconsider the approaches for READ and WRITE operations, suggested in [1]–[5],[7], using MATLAB and SPICE simulations. A novel write technique has been realized here with 3 FinFET structure using a single word line and read operation with a single read line. The developed approach considers the general case of asymmetric MTJ [6]. In this paper, we have shown the Spice simulation result of MTJ for optimizing the peripheral devices and have the effects of peripheral devices on the MTJ operations.

## II. Write Operation

The write circuit and the pulses has been depicted in Fig 1 has been explained in the Fig 2. The word line is connected to an inverter circuit as shown in the Fig.1. The suggested writing scheme is it contains only one line, as the main advantage, which makes it suitable for crossbar memory array implementation. We have adopted the BSIM-CMG [8] and ASU MRAM model [7] to predict the switching features with varying gate length ($L_g$) and width ($T_{Fin}$). Fig 3 and Fig 4 represent AP to P switching where, we found the best combination is $T_{Fin}$=10nm and $L_g$=20nm. The switching delay and $I_{ON}$, $I_{OFF}$ ratio in Fig 5 proves our conclusion. The propagation delay in '0' to '1' bit state is 3.5ns, while it takes 3ns for the reverse operation, where the delay in [7] was 5ns and 7ns. Fig 5 and Fig 8 shows switching performance of MTJ with varying dimensions of the FinFET. Here we have tried to optimize the peripheral device dimensions for MTJ and found that FinFET with dimensions $T_{Fin}$=10nm and $L_g$=20nm are the most viable combination in terms of power consumptions and delay.

## III. Read Operation

To produce a random binary code a full cycle of the switching can be reduced down to two sub-cycles, Fig 10. The MTJ starts at AP state (a) and can be switched or not at (b); at phase (c) the signal becomes detected after the *write* cycle with the probability P ≈ 0.5. During the *read* and *reset* cycle a reverse current is applied, which causes the state either to switch back from P to AP or let it stay in AP. In the end of the cycle the state will be *reset* to AP for the next *write* cycle, phase (d) in Fig 9. Along the *read* and *reset* sub-cycles, the voltage of the represented voltage divider of the MTJ-cell with "Read and Reset" FINFETs is determined at a measuring point by the full bridge, Fig 9. The measured voltage ($V_m$) depends on the state of the MTJ at the sensing pin.. In case of the MTJ is pinned at the *read* and *reset* cycle from P to AP, a voltage reduction at the measuring point is observed and detected in similar way as in [3]. This voltage deviation is amplified by an inverting comparator. The signal output of the comparator is detected by the delay flip-flop (DFF): P to AP switching can be indicated as digital "1", otherwise, the output of the DFF determined as digital "0".

## IV. Conclusion

As a result, in this work, the concept of TRNG with two sub-cycles per period have been developed. The optimized MTJ-based model of TRNG includes four FinFETs connected as a full bridge *via* MTJ cell. Calibration procedure is suggested to tune the amplitude of the current pulse, which have to provide the equal probability for both MTJ states taking into account the MTJ asymmetry or signal deviations within lifetime. The model can be used in applications for a computational memory concepts, providing longer electronic lifetime. One of the benefits of the developed technique is the increased amount of true random numbers per time by the cause of restricted number of logical cycles.

## References

[1] "Implementing p-bits with embedded MTJ," *S.Dutta et al*.
[2] "A novel circuit design of true random number generator using magnetic tun- nel junction,"Y.Wang et al.
[3] Y. Qu et al "A true random number generator based on parallel STT-MTJs,"
[4] "A magnetic tunnel junction based true random number generator with conditional perturb and real-time output probability tracking," W. Ho Choi et al.
[5] I. Chakraborty, A. Agrawal, K. Roy, "Design of a low-voltage analog- to-digital converter using voltage-controlled stochastic switching of low barrier nanomagnets," *IEEE Magn. Lett*., vol. 9, 3103905, May 2018.
[6] A. Useinov, and J. Kosel, "Spin asymmetry calculations of the *TMR- V* curves in single and double-barrier magnetic tunnel junctions," *IEEE Trans. on Magn*., vol. 47, no. 10, pp. 2724-2727, Oct. 2011.
[7] Z. Xu, C. Yang, M. Mao, K. B. Sutaria, C. Chakrabarti, Y. Cao, "Compact modeling of STT-MTJ devices," Solid-State Electronics, vol. 102, pp. 76-81, December 2014.
[8] "A Multi-Gate CMOS compact model-BSIM-MG" Darsen Lu et al.

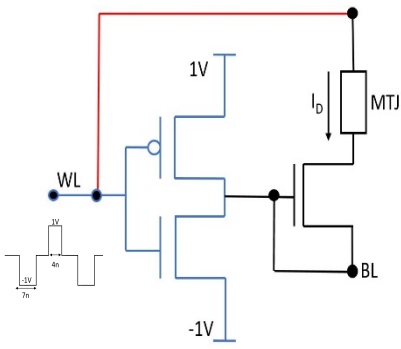
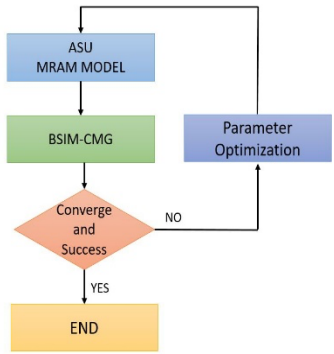
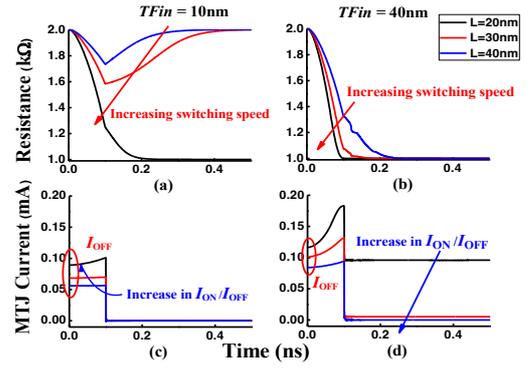

**Fig 1:** Write Circuit for MTJ

**Fig 2:** Simulation Setup

**Fig 3:** Anti-Parallel to Parallel Transition-I
(Switching dynamics variation with $L_g$)

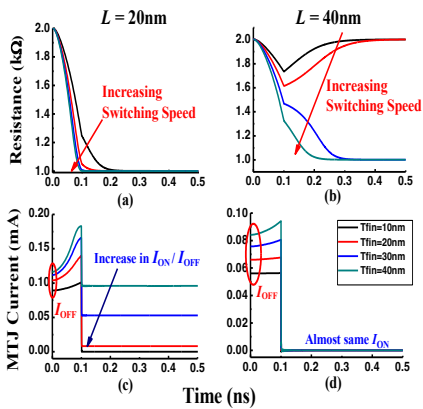
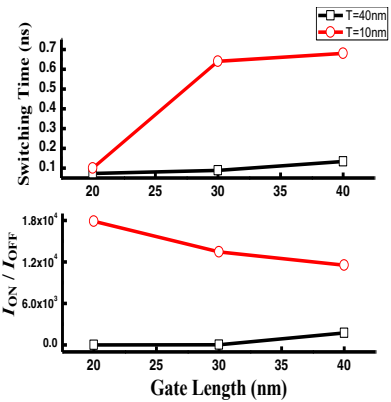
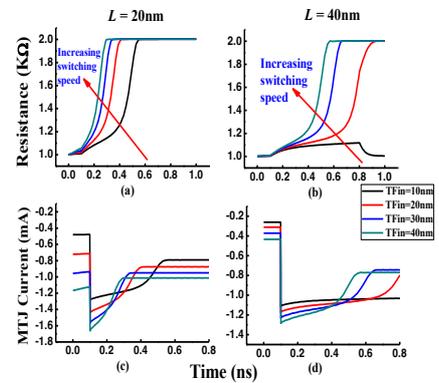

**Fig 4:** Anti-Parallel to Parallel Transition-II
(Switching dynamics variation with $T_{Fin}$)

**Fig 5:** Delay and $I_{ON}/I_{OFF}$ vs $L_g$

**Fig 6:** Parallel to Anti-Parallel Transition-I
(Switching dynamics variation with $T_{Fin}$)

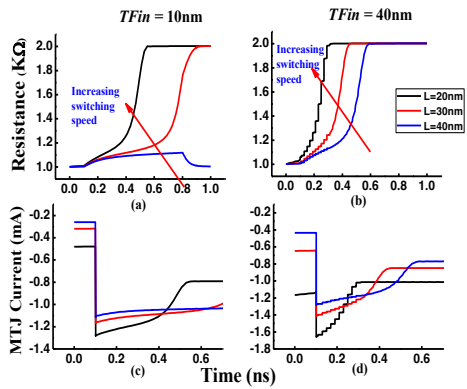
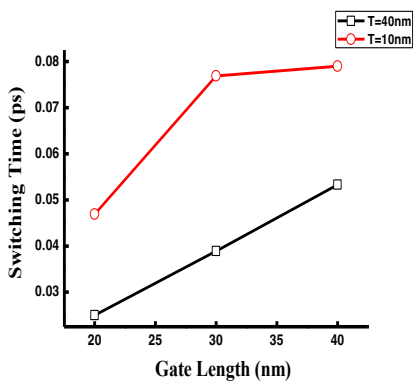
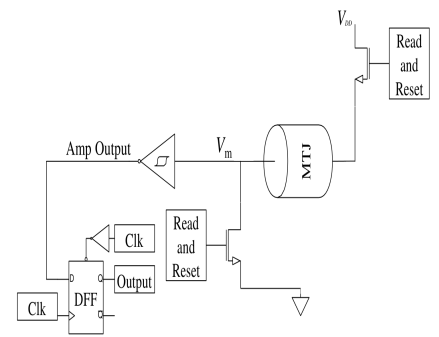

**Fig 7:** Parallel to Anti-Parallel Transition-II
(Switching dynamics variation with $L_g$)

**Fig 8:** Delay vs $L_g$

**Fig 9:** Read Circuit: MTJ based model of the computational block of TRNG. The signal direction is taken from right to left. DFF signal represents the output which generate p-bits.

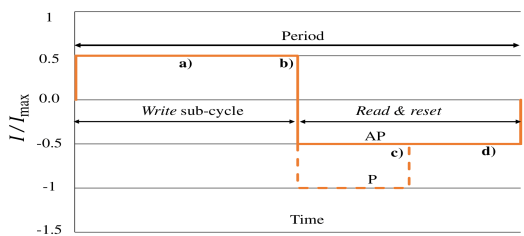

**Fig 10:** Current through the MTJ during the whole period.